\documentclass[a4paper,twocolumn,showpacs,amsmath,pra,amssymb]{revtex4-1} %letterpaper,preprint
\usepackage[colorlinks=true,urlcolor=blue,citecolor=blue,linkcolor=blue]{hyperref}

\usepackage{amssymb}
\usepackage[mathcal]{euscript}
\usepackage{bm}
\usepackage{parskip}

\usepackage{times}
\usepackage{color}
\usepackage{xspace}
\usepackage{amssymb,amsmath,amsfonts}
\usepackage{amsbsy}
\usepackage{graphicx}
\usepackage{epstopdf}
\usepackage{float}
\usepackage{pslatex}
\usepackage{contour}
\usepackage{color}
\usepackage[usenames,dvipsnames]{xcolor}
\newcommand{\ket}[1]{|#1\rangle}
\newcommand{\etal}{{\it et al.\/}\ }
\newcommand{\eref}[1]{Eq.~(\ref{#1})}
\newcommand{\sref}[1]{Sec.~\ref{#1}}
\newcommand{\fref}[1]{Fig.~\ref{#1}}

\newcommand{\Sref}[1]{Section~\ref{#1}}
\newcommand{\Fref}[1]{Figure~\ref{#1}}

\definecolor{raw}{RGB}{255,177,100}
\definecolor{filtered}{RGB}{153,0,51}
\begin{document}
\title{Dispersive probing of driven pseudo-spin dynamics in a gradient field}
\author{Amita B. Deb}
\author{Bianca J. Sawyer}
\author{Niels Kj{\ae}rgaard}\email{nk@otago.ac.nz}
\affiliation{Jack Dodd Centre for Quantum Technology, Department of Physics, University of Otago, Dunedin, New Zealand.}
\date{\today}
\begin{abstract}
We have studied the coherent evolution of ultracold atomic rubidium clouds subjected to a microwave field driving
Rabi oscillations between the stretched states of the $F=1$ and $F=2$ hyperfine levels. A phase winding of the two-level system
 pseudo-spin vector is encountered for elongated samples of atoms exposed to an axial magnetic field gradient and
can be observed directly in state-selective absorption imaging. When dispersively recording the sample-integrated spin
population during the Rabi drive, we observe a damped oscillation directly related to the magnetic field gradient,
which we quantify using a simple dephasing model. By analyzing such dispersively acquired data from millimeter sized
atomic samples, we demonstrate that field gradients can be determined with an accuracy of $\sim25$~nT/mm. The dispersive probing
of inhomogeneously broadened Rabi oscillations in prolate samples opens up a path to gradiometry with bandwidths in the kilohertz domain.
\end{abstract}
\maketitle
\section{Introduction}
Two-level systems are ubiquitous in quantum physics
and have been the subject of theoretical studies for almost a century \cite{Zener1932,Rabi1937,Barnes2012}. They constitute the fundamental building blocks for quantum information processing \cite{Zoller2005} and time keeping with atomic clocks \cite{Audoin2001,Riehle2004}.
For such quantum enabled technologies, physical systems that display a high degree of immunity with respect to
the surrounding environment are invariably sought. A prominent example is optical lattice clocks \cite{Derevianko2011}, where magnetically insensitive ``clock-state'' atoms are confined by ``magic'' wavelength optical fields. The addition of coupling terms (interaction with the environment) to the free evolution Hamiltonian for a two-level system can, however, be turned to an advantage and exploited in metrology applications.
In this case, a field induced level shift is encountered, which may be measured spectroscopically. This may form the basis of atomic magnetometers \cite{Dupont-Roc1969,Kominis2003,Fatemi2010} and magnetic field imaging \cite{Terraciano2008a}.
\setlength{\parskip}{3pt}

Two-level atoms are typically manipulated with quasi-resonant radiation fields, giving rise to dynamics equivalent to a magnetic spin-1/2 particle in a B-field \cite{Feynman1957}. The physics of a two-level atom in a steady field is captured by the well-known Rabi solution describing the oscillation of population between the two states, which for an atom initially in its lower energy state takes the following form for the (single atom) population inversion
\cite{Allen1975}
\begin{equation}\label{eqrabi}
w(t, \Delta)=-1+2\left[\frac{\chi}{\Omega(\Delta)}\right]^2\sin^2\left[\Omega(\Delta)t/2\right],
\end{equation}
where $\chi$ is the on-resonance Rabi frequency and
\begin{equation}\label{eqrabifreq}
\Omega(\Delta)=\sqrt{\chi^2+\Delta^2},
\end{equation}
is the generalized Rabi frequency for a field detuned by $\Delta$ from the transition frequency. The on-resonance Rabi frequency is the product of the driving field amplitude and the matrix element between the two states for the atom-field interaction, and may hence display spatial variation according to the intensity distribution of the electromagnetic field. This was recently exploited to map out the magnetic microwave near-field of a coplanar waveguide by means of ultracold \cite{Bohi2010} and hot \cite{Bohi2012} atomic clouds. Similarly, in the far field of a horn antenna, the effect of a spatially inhomogeneous microwave drive field has also been observed for extended atomic clouds \cite{Daniel2013}.
The dynamics of the inversion $w(t,\Delta)$ may also acquire a position dependence through a spatial variation of the detuning $\Delta$. This could, for example, arise from a position dependent differential ac Stark shift imposed by a laser beam \cite{Windpassinger2008a} or via a nonuniform magnetic field (Zeeman shift) \cite{Dupont-Roc1969,Kominis2003,Fatemi2010,Terraciano2008a}.

In this paper we report on dispersive probing of microwave driven Rabi oscillations between two Zeeman tunable hyperfine states in a prolate sample of ultracold $^{87}$Rb atoms. We organize our presentation as follows. In \sref{section:background} we briefly review dispersive probing schemes and their applications, and discuss approaches for imprinting magnetic order in inhomogeneously broadened samples. \Sref{section:setup} describes our experimental setup and the acquisition and processing of data. For our atomic system, a magnetic field gradient over an elongated (prolate) cloud leads to an inhomogeneous shift in the resonance. This in turn gives rise to a decay in Rabi oscillations when probed dispersively along its axis. \Sref{section:interplay} presents examples of such decaying waveforms, which are interpreted in a simple dephasing model and compared to complementary data acquired by means of spatially resolved state-selective absorption imaging. Here we also discuss the small but observable effect the dispersive probe beam has on the atomic sample in terms of a differential light shift, breaking the symmetry for the resulting spin texture. We finally summarize our findings and outline potential future directions in \sref{section:conclusion}.

\section{Background}
\label{section:background}
\subsection{Dispersive probing}
When a beam of light of well-defined frequency passes through a gas of atoms it may be attenuated as a result of absorption, as well as shifted in phase.
Dispersive detection techniques make use of the latter effect and are a very suitable tool for tracking dynamical processes in ``real time'' for cold and ultracold atomic systems. For example, dispersive probing has been used to follow breathing \cite{Petrov2007} and center-of-mass oscillations \cite{Kohnen2011}, and the process of forced evaporative cooling \cite{Sawyer2012} for trapped atomic clouds. Moreover, coherent phenomena such as Rabi oscillations between hyperfine states \cite{Chaudhury2006,Windpassinger2008,Bernon2011}, Larmor precession \cite{Isayama1999} and phase space dynamics of spinor condensates \cite{Liu2009} have been recorded and studied dispersively. The phase shift measurement is obtained by comparing the phase-shifted beam to a reference beam in either homodyne or heterodyne detection schemes. For example, the former is achieved in Mach-Zehnder interferometers \cite{Petrov2007}, while the latter is achieved in frequency modulation (FM) spectroscopy \cite{Lye1999}. Furthermore, dispersive probing may also take the form of polarization spectroscopy for gases in birefringent states \cite{Isayama1999,Chaudhury2006}.

The implementations listed in the preceding paragraph all make use of non-imaging photodetectors, which are generally both faster and less expensive than low noise CCD cameras \cite{Lye1999}, but do,
of course, imply that direct spatial information imprinted on the probe light is integrated out. Spatially resolved dispersive imaging techniques for ultracold atoms have been in play since right after the achievement of Bose-Einstein condensates (BECs) \cite{Andrews1996}, with variants thereof developing subsequently \cite{Andrews1997,Bradley1997,Kaminski2012,Gajdacz2013}.
\subsection{Imprinting continuous magnetic order in a prolate atomic sample}\label{section:backgroundb}
Systems with magnetic order are of fundamental importance in condensed-matter physics and an area of substantial interest. Trapped ultracold atomic gases provide a clean and highly flexible experimental test bed where to simulate such systems and have been utilized in many studies over the last 15 years or so to investigate spatial magnetic order by using atomic spin degrees of freedom. The starting point for many such experimental studies in a continuous geometry is to expose the gas to a spatially varying coupling field, inhomogeneously driving dynamics between spin states. For example, Matthews \etal \cite{Matthews1999a} used a spatially varying Rabi drive to wind up the phase of the order parameter of a two-component $\rm^{87}Rb$ condensate and observed the subsequent unwinding owing to its kinetic energy. The spatial variation of the coupling resulted as a combined effect of the presence of a magnetic field gradient and the accompanying variation of the two-photon Rabi frequency. Similar methods were used later to excite dipole topological modes in condensates \cite{Williams1999}, to create spin waves in an ultracold gas above the BEC transition \cite{McGuirk2002,McGuirk2010}, to observe spatial segregation in both ultracold Bose gases \cite{Lewandowski2002} and Fermi gases \cite{Du2008}. Very recently, phase-winding in an elongated two-component BEC was used to generate dark-bright solitons \cite{Hamner2013}. The phenomenon of spin waves has direct relevance to spintronics and spin caloritronics \cite{Wong2012}, and studies have demonstrated universal spin transport in Fermi gases near the hydrodynamic regime \cite{Sommer2011, Koschorreck2013} and engineered high-quantum number spin waves \cite{Heinze2013}. Spatially inhomogeneous coupling has been employed in \cite{Maineult2012} to induce spin waves and study the resulting shift of the clock frequency in an atom chip experiment. In relation to this, there has also been substantial interest in studying two-component BECs, where spatio-temporal variation of Rabi coupling was employed in association with controlling the atomic interactions to study spatial mixing-demixing dynamics \cite{Stenger1998, Hall1998, Nicklas2011}, spin textures following a quench \cite{Sadler2006} and non-equilibrium dynamics \cite{Mertes2007}.

In the present paper, we apply a phase-winding technique similar to \cite{Matthews1999a,Hamner2013} for imprinting magnetic order in an elongated sample of ultracold atoms. The overall magnetization $\overrightarrow{\mathcal{M}}$ of the ensemble exhibits itself as a density weighted integral of the local magnetization, the latter containing information about the local coupling strength. The temporal evolution of $\overrightarrow{\mathcal{M}}$ has a characteristic non-exponential profile, which we map out using both dispersive and absorptive probing. From this, the variation in local coupling strength can be inferred, making the system a prospect tool for high bandwidth magnetic and microwave field gradiometry.

\section{Experimental}\label{section:setup}
\begin{figure*}[tb!]
\begin{center}
\includegraphics[width=0.9\textwidth]{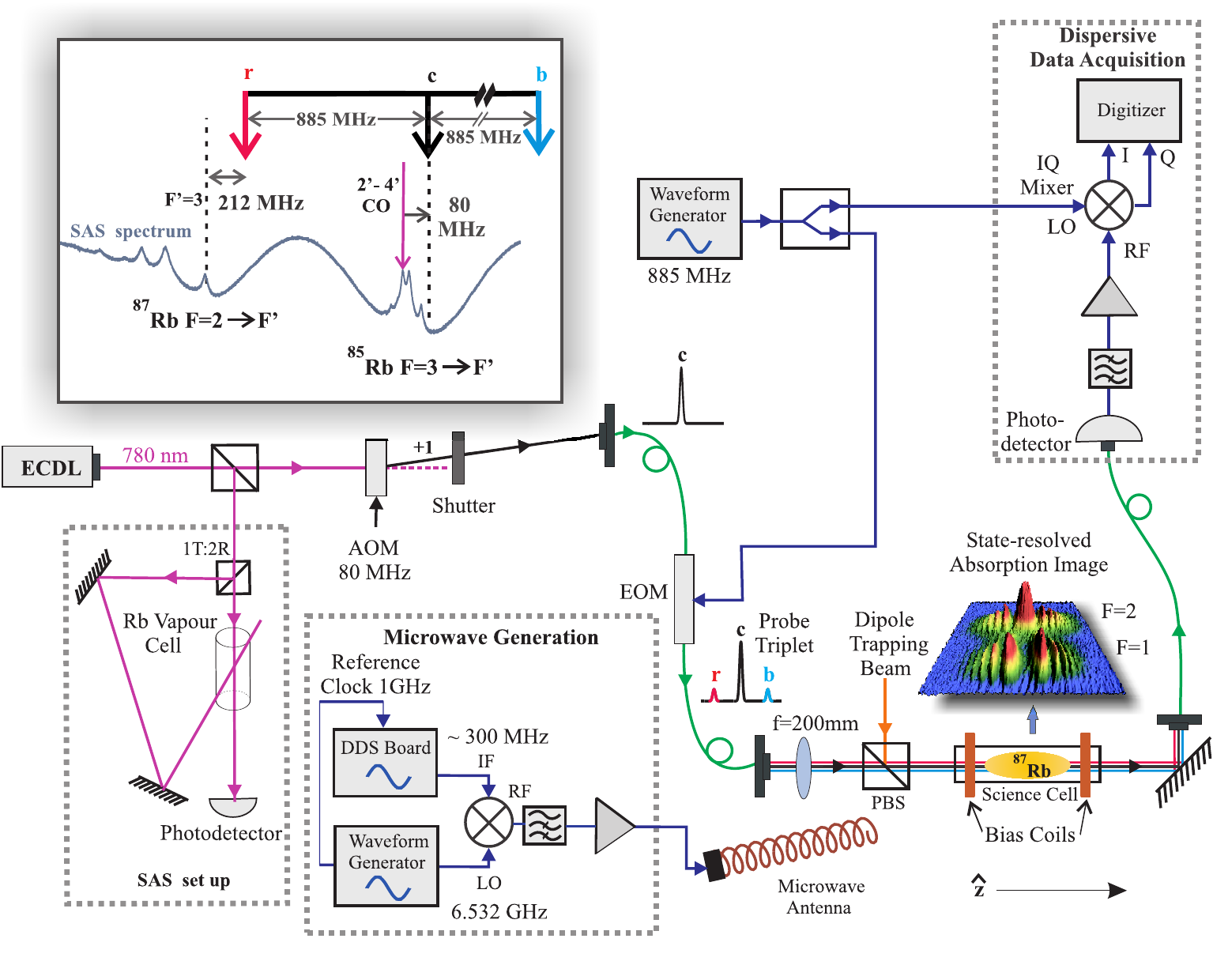}
\caption{(Color online) Experimental setup. A dispersive probe beam is frequency locked via a SAS set up, passed through an AOM for
switching, and an EOM to produce frequency sidebands. The probe pulse frequency triplet then propagates through the
dipole trapped $^{87}$Rb sample in the science cell, and the resulting signal is detected on a fast ac photodetector.
The detector output is demodulated and sampled with a digitizer to recover the optical phase shift information. Inset: The frequencies of the components of the probing beam, relative to the $^{85}$Rb and $^{87}$Rb D2 lines, are indicated.}
\label{fig:SetUp}
\end{center}
\end{figure*}

\subsection{Setup}
\label{section:setupa}
Our experimental setup is shown schematically in \fref{fig:SetUp}. The ultracold atomic samples used in our experiments typically consist of $3\times10^6$ $^{87}$Rb atoms in the $\ket{F=2,m_F=2}$ ground hyperfine state, confined in a single beam far-off-resonant dipole trap \cite{Grimm2000}. Samples are produced using a standard BEC apparatus \cite{Sawyer2012}. Atoms are initially magnetically confined in a Ioffe Pritchard (IP) trap, where they are rf evaporatively cooled to temperatures $<1$\,$\mu\rm{K}$, close to the BEC transition. The sample is then transferred to the dipole trapping potential, produced by a 1064\,nm Ytterbium fiber laser with typical optical power $\sim1.2$\,W and a beam waist of $60$\,$\mu\rm{m}$. The confining beam propagates along the axial ($z$) direction of the cloud, and gives rise to radial and axial trapping frequencies of $(\omega_{\rm{r}}, \omega_{\rm{z}})=2\pi \cdot (270, 1)~\rm{Hz}$. The initial peak atomic density in the dipole trap is $4\times 10^{12}~$cm$^{-3}$.

A magnetic bias field ($\sim 2~\rm{G}$) pointing in the $z$-direction lifts the degeneracy between the Zeeman states and defines a quantization axis. The field is generated with a coil pair symmetrically placed around the cloud in an approximate Helmholtz configuration. The axial field produced by the geometry is well-approximated near the center to be linear with position with a gradient $\sim$ 50~\,mGA$^{-1}$cm$^{-1}$. While the bias field is present, a microwave coupling field, near-resonant with the $|F~=~2,~m_F~=~2\rangle \leftrightarrow |F = 1,~m_F = 1\rangle$ $^{87}$Rb hyperfine transition, is applied to the sample for a variable duration $t_{\rm{mw}}$. This microwave field is generated by mixing of signals from a waveform generator (6.532~GHz) and a direct digital synthesizer (DDS) board ($\sim 300~\rm{MHz}$) using a double-balanced frequency mixer. The mixer output is high-pass filtered to give the up-shifted (sum) component, which is amplified to a power of 6.5~W. A helical end-fire antenna \cite{Balanis1996} emits circularly polarized microwave radiation towards the atomic cloud. The antenna is positioned $\sim 7~\rm{cm}$ from the atoms and, due to space constraints, at an angle of $10^{\circ}$ with the $z$-axis. This configuration results in a predominantly circularly polarized microwave radiation \footnote{We measured the ratio of coupling strengths for $\sigma^{+}$ and $\sigma^{-}$ transitions spectroscopically to be $\sim 10:1$ for this antenna orientation (taking into account the magnetic dipole matrix elements).}. The variation of microwave power over the extent of the atomic cloud is less than 2\%, negligible for the purposes of the studies presented below.

The emitted microwave field induces Rabi oscillations between the $\ket{F=2,m_F = 2}$ and $\ket{F=1,m_F=1}$ states with a characteristic on-resonance Rabi frequency of $\chi \sim 2\pi\times13.5$~\,kHz. We monitor the evolution of the atomic $F=2$ population of the sample during the microwave pulse dispersively, using FM spectroscopy \cite{Bjorklund1980,Lye1999}. To produce an optical frequency triplet for probing, a central carrier (c), far from any optical transitions in $^{87}$Rb, is derived from an external cavity diode laser (ECDL), which is locked to the ~$2^{'}$-$4^{'}$ crossover peak of the $F=3 \rightarrow F'$ D2 transition of the $^{85}$Rb isotope by means of a saturated absorption spectroscopy (SAS) setup (see inset panel of \fref{fig:SetUp}). An acousto-optic modulator (AOM), used to pulse the probe beam on and off, shifts the carrier up by 80\,MHz, such that it is blue-detuned by 1097\,MHz from the $^{87}$Rb $\ket{F=2} \rightarrow \ket{F'=3}$ transition. The carrier is then passed through a fiber electro-optic modulator (EOM, bandwidth 10~\,GHz), which imprints a red (r) and a blue (b) sideband shifted by $f_{\rm{EOM}}=885$~MHz from the carrier frequency, such that the red sideband is blue-detuned by 212~MHz from the $^{87}$Rb $\ket{F=2} \rightarrow \ket{F'=3}$ transition. The three frequency components have approximately equal optical power ($\sim 1.7~\mu\rm{W}$ each).

A switching routine is administered by the AOM, in which a sequence $i = 1,...,N$ (with $N$ typically 50~--~300) of 500~ns long probe light pulses is produced at a repetition rate of 100\,kHz. The probe beam is focused to a waist of 65\,$\mu$m onto the atoms such that the frequency triplet propagates along the axis of our atomic sample. The probe beam is linearly polarized, and can be decomposed into right and left circularly polarized components of equal intensity, which address the $\ket{F=2,m_F=2} \overset{\sigma^+}{\rightarrow} \ket{F'=3,m_F=3}$ and $\ket{F=2,m_F=2} \overset{\sigma^-}{\rightarrow} \ket{F'=1,2,3,m_F=1}$ transitions, respectively. Atom-light interactions cause each frequency ($q_1 = c, r, b$) and polarization ($q_2 = +, -$) component to acquire a phase shift $\phi_{q_1}^{q_2}$. For the probe beam parameters and maximum atomic column densities in the $F = 2$ state specified above, we estimate $\phi^{+}_r=4.76^{\circ}$, $\phi^{+}_c=0.92^{\circ}$, $\phi^{+}_b=0.51^{\circ}$, $\phi^{-}_r=0.83^{\circ}$, $\phi^{-}_c=0.25^{\circ}$, and $\phi^{-}_b=0.15^{\circ}$. The resulting optical beat signal is detected on a fast, fiber-coupled photodiode (Finisar HDF3180-203, bandwidth 4.2~\,GHz) and is demodulated to baseband with a local oscillator (LO) at $f_{\mathrm{EOM}}$ using an IQ mixer. In this way, components both in-phase (I) and in-quadrature (Q) with the LO are obtained, so the signal can be reconstructed irrespective of the LO phase. The signals from the two mixer output ports are simultaneously sampled at 20~MS/s by a 16-bit digitizer in $5~\mu\rm{s}$ segments centered on each pulse. The single-pulse signal-to-noise ratio of the dispersive signal for the maximum atomic numbers we use is about 20.

Our experimental sequence is (crucially) triggered from the rising edge of a square wave signal synchronous to the 50~Hz power line frequency. A suitable delay is chosen to ensure that the Rabi drive and dispersive probing happen around a zero crossing for the residual ac magnetic field originating from power line currents \cite{Schmidt-Kaler2003,Terraciano2008}. The timing of the microwave and the dispersive probe pulses is controlled by digital delay generators, such that similar experimental conditions are realized in every experimental run.

The cloud may be additionally (or alternatively) probed using conventional absorption imaging at the end of the experimental sequence, after switching off the dipole trap confinement. Atoms falling under gravity are spatially separated in the vertical direction based on their magnetic moment by a Stern-Gerlach gradient of $\sim 12~\mathrm{G/cm}$, applied for $\sim7~\mathrm{ms}$. Atoms are then exposed to an optical ``repumping'' pulse, which transfers $F=1$ atoms to $\rm F=2$ via the $^2P_{3/2}$ ($F'=2$) excited state, and finally imaged with probe light resonant with the $\ket{F=2} \rightarrow \ket{F'=3}$ transition. This procedure produces an absorption image in which the spatial distributions of the two hyperfine states originally present are clearly distinguishable (see \fref{fig:SetUp}).

\subsection{Data acquisition and analysis}
\begin{figure}[tb!]
%\begin{center}
\flushright
  \includegraphics[width=\columnwidth]{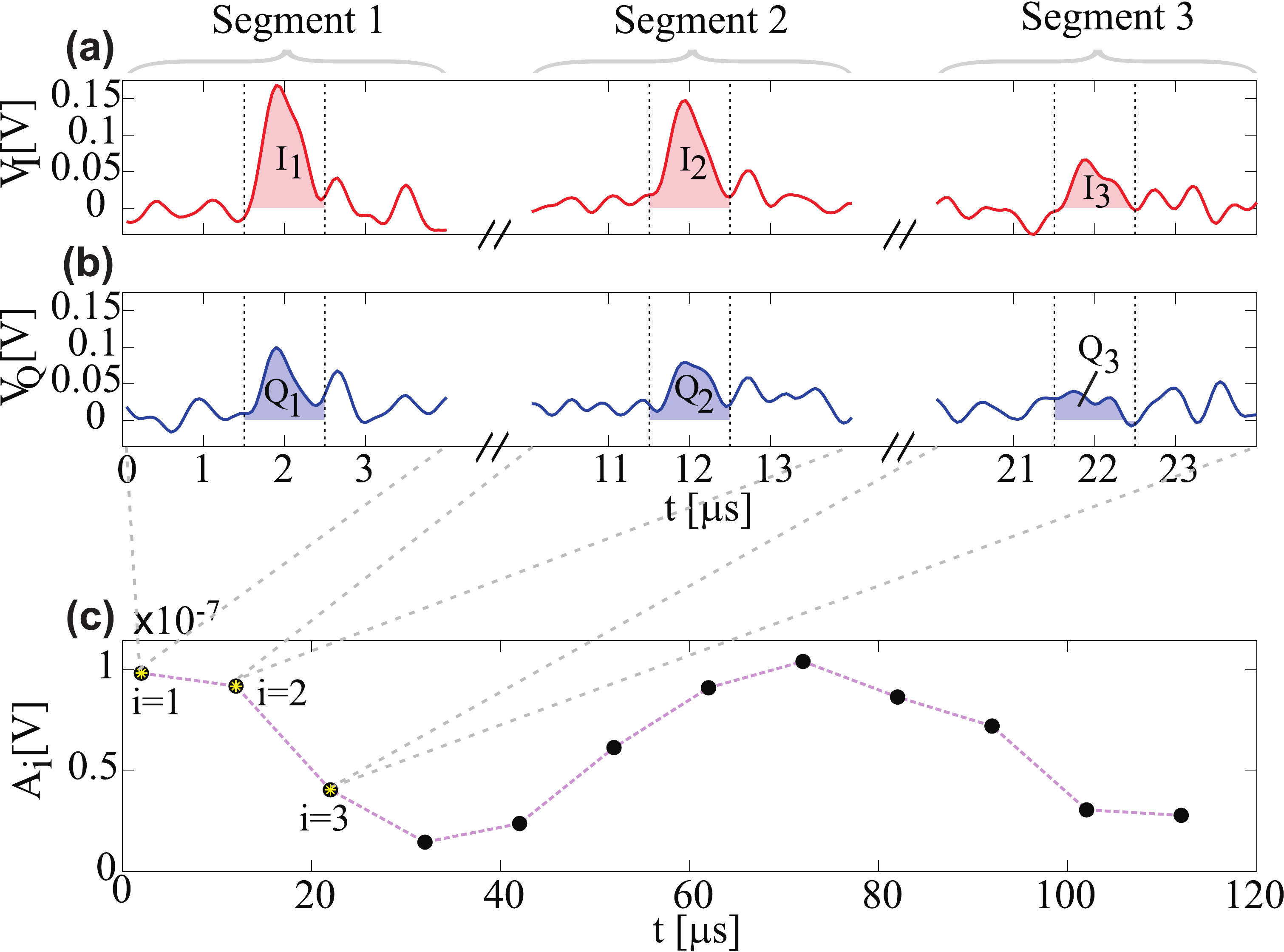}\\
  \caption{(Color online) Schematic representation of the data processing procedure for a typical data set. The upper two panels show the raw data collected on each channel of the digitizer for $i = 1,2,3$. The dotted black lines indicate the integration window for each sample (note the broken axis). The lower panel shows the resulting phase shift data for this data set.}\label{fig:PulseAnalysis}
%\end{center}
\end{figure}
The raw data for probe pulse $i=1,...,N$ acquired by the digitizer consists of $V_I^i$ and $V_Q^i$ --- the voltages obtained from the I and Q ports of the mixer, respectively. Within each data segment we define an integration window and numerically extract the area under the probe pulse to give the processed data sets $\{I_i,Q_i\}$. This integration is shown schematically in \fref{fig:PulseAnalysis}\,(a) and \fref{fig:PulseAnalysis}\,(b), for the first three points of a typical data set showing Rabi oscillations.

The phase shift $\phi_r^+$ incurred by the red probe sideband, assumed to have a low spontaneous scattering probability, can be shown to be proportional to $A_i~=~\sqrt{I_i^2+Q_i^2}$ for $\phi_r^+ \ll \pi$, which is true under our conditions. The resulting dispersive signal, which is proportional to population of atoms in the $F=2$ state, is plotted in \fref{fig:PulseAnalysis}\,(c), with the values corresponding to the raw data shown above indicted with yellow stars.

\section{Interplay between field gradient, magnetic order and dispersive probing}
\label{section:interplay}

\subsection{Dispersive monitoring of Rabi oscillations}
\label{section:disprabi}
As described in section \ref{section:setupa}, our starting point for the coherent state manipulation is an ultracold sample polarized in state $|F=2,m_F=2\rangle$, loaded in a focused single-beam Gaussian dipole trap. The depth of the dipole trap is $\sim 25$\,$\mu$K, such that we achieve a loading efficiency close to unity. The small atomic sample expands along the direction of the dipole trap beam.
After 40~ms expansion, for example, the cloud has a FWHM size of 1.3~mm in the axial direction and about 60~$\mu$m along the radial direction, such that we have a quasi-1D prolate cloud of atoms. The typical shot-to-shot number and temperature stability of the preparation of the initial atomic sample is better than 5\%.
\begin{figure}[tb!]
\flushright
   \includegraphics[width=\columnwidth]{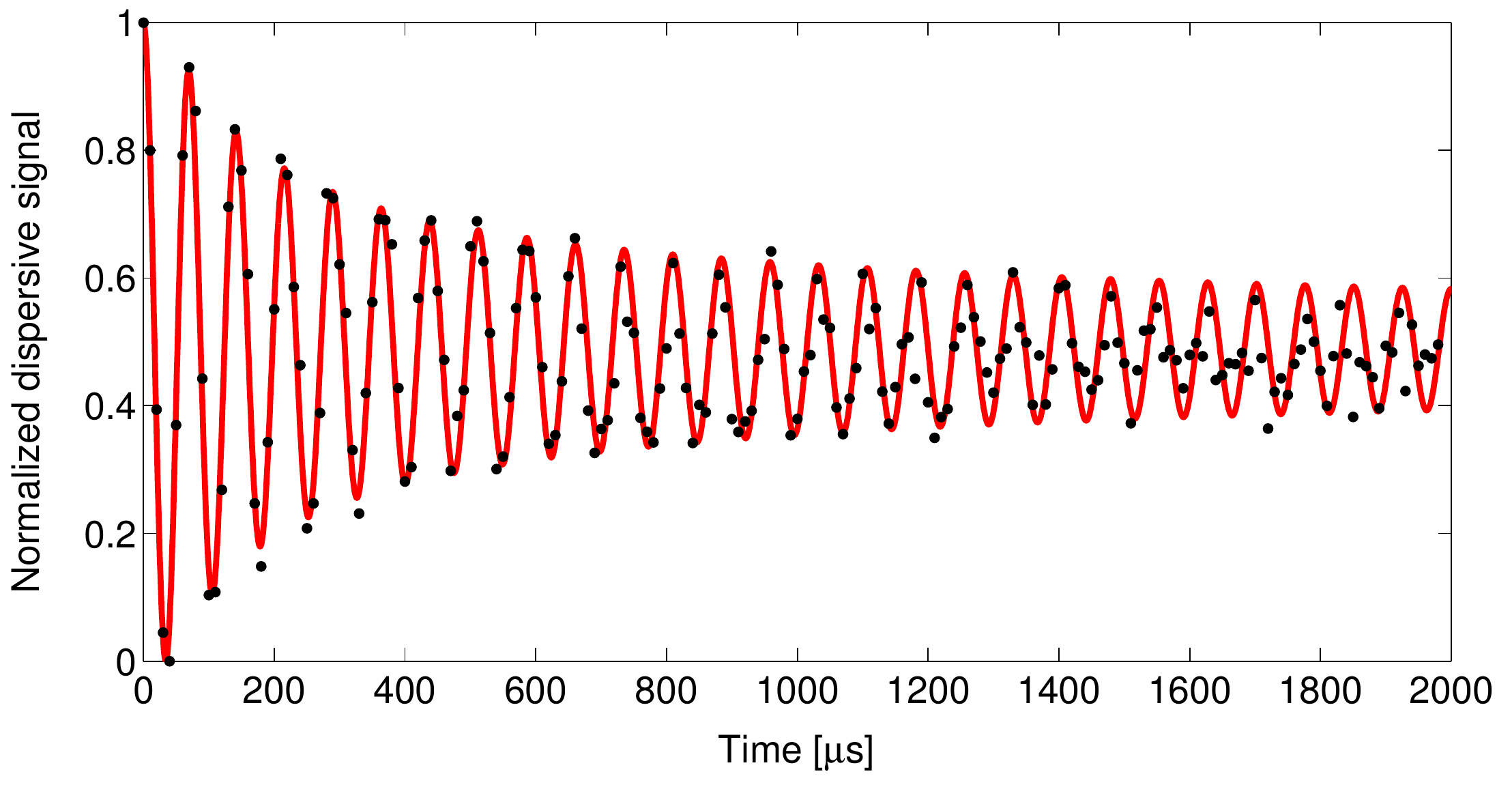}\\
  \caption{(Color online) Dispersive monitoring of Rabi oscillations in a quasi-1D atomic cloud ($\bullet$). The plot is an average of five consecutive experimental runs. The atomic sample was allowed to expand for 40~ms in the horizontal trap before applying the Rabi drive and the dispersive pulses. The line is a fit of the model described in \sref{section:gradient_magnetimetry} to our data.}\label{fig:disprecord}
\end{figure}

In \fref{fig:disprecord}, we show a typical record of the dispersive interrogation of Rabi oscillations of atoms driven by microwave for 2~ms. The expansion time of the atoms in dipole trap was 40~ms prior to microwave exposure, and the data points are averages of 5 consecutive experimental runs. The decay of the oscillation is predominantly a result of magnetic dephasing of the Zeeman-sensitive states (which leads to the formation of spin domains, as discussed in \sref{section:spinwave}).
\subsection{Absorptive measurements: formation of magnetic order during Rabi flopping}\label{section:spinwave}
\begin{figure}[tb!]
\flushright
\includegraphics[width=\columnwidth]{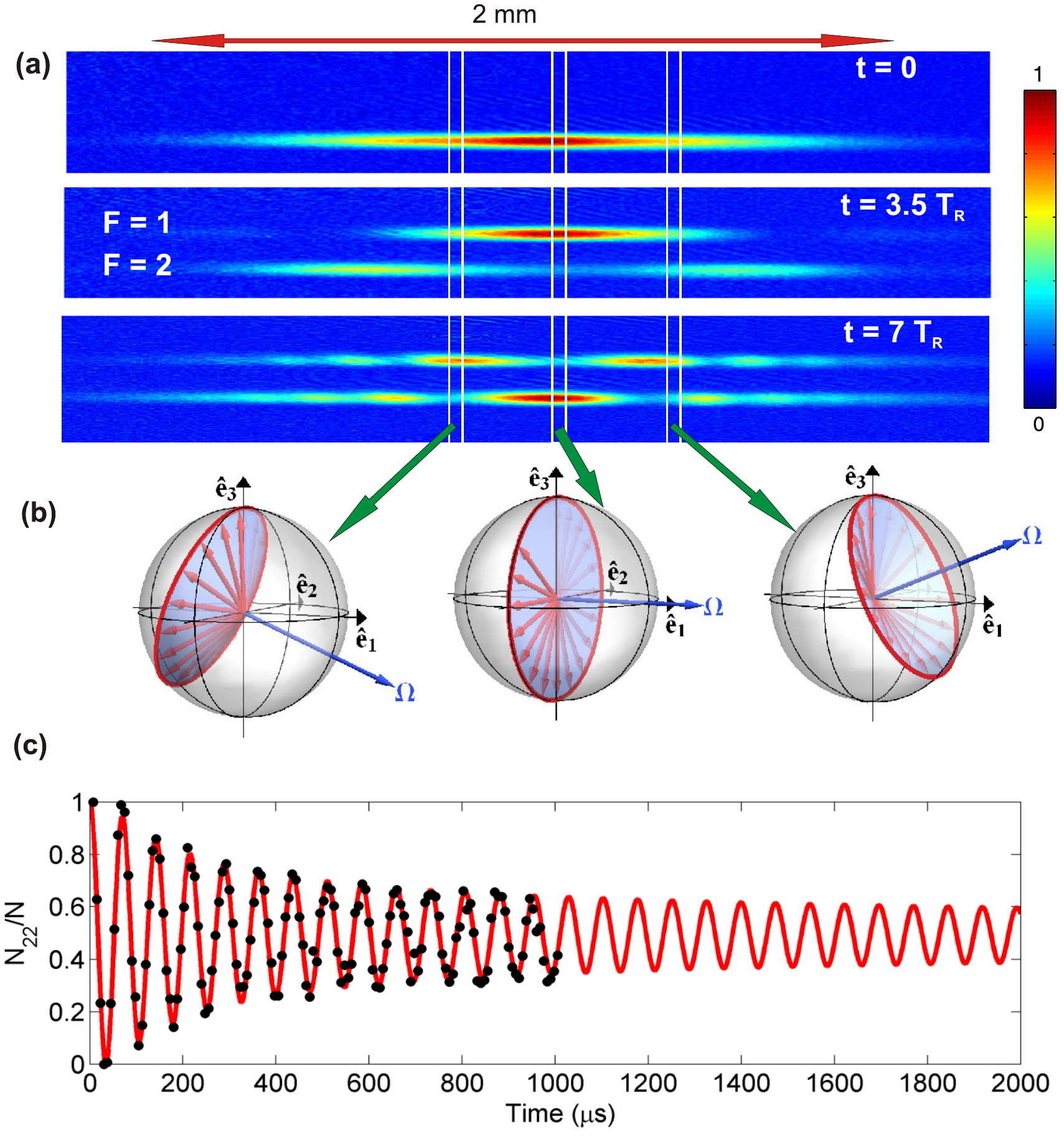}
  \caption{(Color online) Emergence of magnetic order in quasi-1D atomic samples. (a) State-resolved ($F=1,2$) absorption images acquired at times $t=\{0, 3.5, 7\}\times T_R$ during the coherent microwave manipulation. (b) The precession of the Bloch vector at the three positions of the atomic cloud shown in (a). The left- and the right-hand side cases are off-resonant, and precess at faster rates. (c) Time evolution of the sample-integrated relative population of atoms in the $F=2$ state from absorption images ($\bullet$). The line is a fit of the model described in section \ref{section:gradient_magnetimetry} to our data. The destructive measurements based on absorption imaging are time-consuming, so we limit ourselves to recording up to 1~ms of microwave manipulation.}\label{fig:bloch1}
\end{figure}

We complement our dispersive data presented in section \ref{section:disprabi} with spatially resolved absorption images acquired at the end of the microwave pulse. We chose the microwave frequency to be resonant with the atoms approximately at the location of the maximum atomic density. \Fref{fig:bloch1}\,(a) shows state-resolved absorption images taken after projective measurements taken at times $t = \{0, 3.5, 7\}\times T_{R}$, where $T_R$ is one full on-resonance Rabi cycle. The emergence of ``spin domains'' is evident, and is due to the spatial variation of Rabi frequency in the presence of a small field gradient.

To elucidate the time evolution of the system, we use a Bloch sphere representation of the two-level system \cite{Foot2005,Allen1975,Riehle2004}. Each point on the $z$-axis is described by an effective pseudo-spin vector $\mathbf{R}(z) = u(z) \hat{\mathbf{e}}_1 + v(z)\hat{\mathbf{e}}_2 + w(z)\hat{\mathbf{e}}_3$ precessing in a plane perpendicular to a torque vector $\bm{\Omega}(z) = \chi\hat{\mathbf{e}}_1 + \Delta(z)\hat{\mathbf{e}}_3$, where $u$ and $v$ are the (local) transverse coherences and $w$ is the (local) population difference between the two states, all defined in a frame rotating at the transition frequency. The equation of motion of the local Bloch vector is $d\mathbf{R}(z)/d t = \mathbf{R}(z) \times \bm{\Omega}(z)$.

\Fref{fig:bloch1}\,(b) illustrates the time evolution of Bloch vectors at three different axial locations in the cloud.  At the center of the cloud, where the microwave is resonant with the atoms, the torque vector is along the $\hat{\mathbf{\mathbf{e}}}_1$ axis, and the Bloch vector precesses on a great circle at a rate $\chi$, whereas away from the center, the Bloch vectors precess on minor circles at a faster rate given by \eref{eqrabifreq}.
Since the phase advance per unit time increases monotonically away from the center of the cloud, this poses a winding mechanism by which alternating domains of the spin projection $w$ are formed.
As an illustration of this dynamics, \cite{movielink} shows the development of the pseudo-spin projection $w$ across the atomic cloud obtained by subtracting the $F=1$ part from the $F=2$ part in a series of state-selective absorption images.

\Fref{fig:bloch1}(c) shows the evolution of the ensemble-integrated population of atoms in the $|F=2,m_F=2\rangle$ state from the absorption data, which exhibits a non-exponential damping of Rabi oscillations mirroring our dispersive observation [\fref{fig:disprecord}]. We note that the time required for the absorptive data acquisition amounted to several hours, in contrast to the dispersive measurements, which only took a few minutes.
\begin{figure}[b!]
%\begin{center}
\flushright
  \includegraphics[width=\columnwidth]{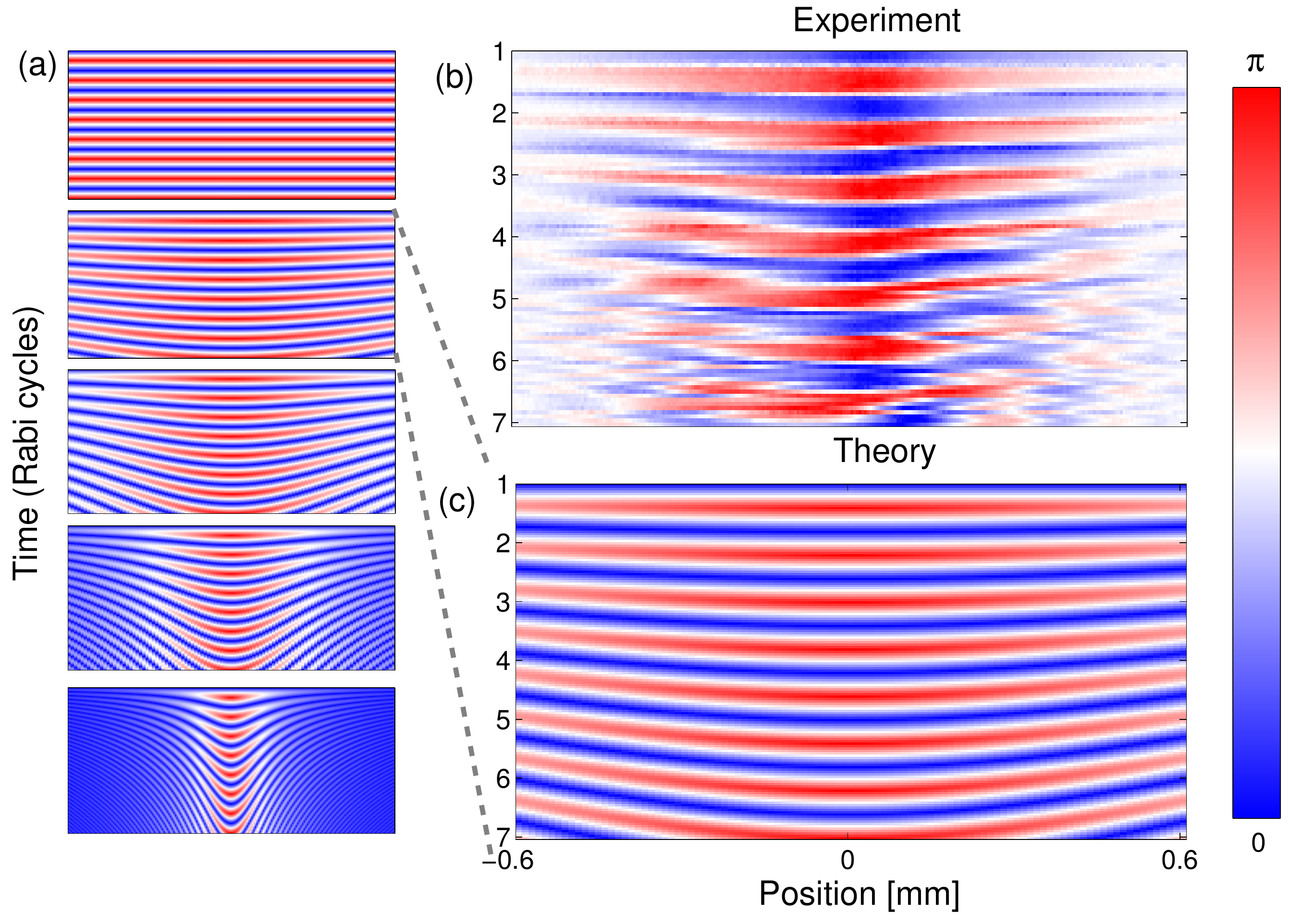}\\
  \caption{(Color online) Maps of the polar angle $\theta (z,t)$ of the pseudo-spin on the Bloch sphere, where $\theta = 0 (\pi)$ corresponds to an atom in $|2,2\rangle$ ($|1,1\rangle$) state. (a) shows the theoretical plots for the cases with $\beta/ \chi = \{0, 1, 2, 4, 10\}~$mm$^{-1}$ shown from top to bottom. (b) shows the map obtained in the experimental sequence. (c) zooms the theoretical map corresponding to $\beta/\chi = 1~$mm$^{-1}$ for easy comparison with the experimental map (see text).}\label{fig:bloch2}
%\end{center}
\end{figure}

\subsection{Dephasing model for an inhomogeneous magnetic field}
As noted in section \ref{section:setupa}, the bias magnetic field in our set up is a linear function of $z$ at the atomic cloud, implying that (via the Zeeman effect) the frequency detuning of the microwave field is of the form $\Delta(z) = \beta z + \gamma$. Since the extent of the cloud is radially small ($\sim 30$~$\mu$m), we assume $\Delta$ to be constant in the radial direction. Furthermore, the effect of mean-field interactions are considered negligible for our experimental conditions as we are in a relatively low density regime and the s-wave scattering lengths for the various spin-states involved are similar to within a few percent. We also neglect any identical spin rotation effect \cite{Lhuillier1982,*Lhuillier1982b} in our system, as the time scale for the exchange collision \cite{Oktel2002} at the densities involved in our experiments is $\sim50$\,ms, which is much longer the than the time scale of the observations made in this paper.

In order to gain quantitative understanding of the ensemble-integrated evolutions presented in \fref{fig:disprecord} and \fref{fig:bloch1}, we numerically solved the equation of motion for $\mathbf{R}(z)$ across the cloud assuming that a linear field gradient along $z$-direction is present and that the center of the cloud is resonant with the microwave field. The maps in \fref{fig:bloch2}(a) show the $z$-dependent time evolution of the ``longitudinal angle''
[i.e., the Bloch vector polar angle $\theta(z,t) = \arccos(w)$] computed for field
gradients $\beta/\chi = \{ 0,1,2,4,10 \} \mathrm{mm}^{-1}$. \Fref{fig:bloch2}(b) shows a corresponding map based on experimentally acquired data. As is evident, the essential features of the phase-winding in our experiments are well-captured by the model based on the presence of a linear field gradient, which suggests a field gradient of around $13.6~\mathrm{kHz~mm}^{-1}$ (\fref{fig:disprecord}(c)).

\Fref{fig:bloch2}(b) shows that after about $4 \times T_R$, the temporal evolution of the atomic pseudo-spin at positions away from the center of the cloud exhibits the emergence of spurious oscillations. The phenomenon cannot be linked to detection noises or fluctuations of magnetic fields, as the center part of the cloud remains unaffected for much longer. At the current stage we have not been able to trace out the underlying mechanism causing this. In the analysis of \emph{local} Rabi frequencies in \sref{section:ex_local_rabi}, we therefore only make use of the first four Rabi cycles.

\subsection{Determining the magnetic field gradient from decaying Rabi oscillations}\label{section:gradient_magnetimetry}
The damping of Rabi oscillations in the ensemble-integrated data obtained through absorption imaging, as shown in \fref{fig:bloch1}(c), and the dispersively acquired data of \fref{fig:disprecord} are both non-exponential and are characterized by a rapid decay in the beginning and a much slower one at later times. Such non-exponential decays are characteristic of spatial non-uniformities, and the two obvious spatially non-uniform features of our system are the local magnetic field and the atomic density. The full model then accounts for the overall pseudo-spin vector of the ensemble as a density-weighted integral of local spins subject to local torque vectors. In particular the population in the $\ket{F=2,m_F=2}$ state at time $t$ is described by,
\begin{equation}\label{model2}
    N_{22}(t) = N_{\mathrm{tot}} - \int d z \overline{n}(z) \frac{\chi^2}{\chi^2 + \Delta^2(z)}\sin^2\left(\frac{\sqrt{\chi^2 +\Delta^2(z)}}{2}t\right),
\end{equation}
where $\overline{n}(z)$ is the initial normalized line density of atoms, which can be inferred from an absorption image taken at $t=0$.
\begin{figure}[bt!]
\begin{center}
  \includegraphics[width=0.8\columnwidth]{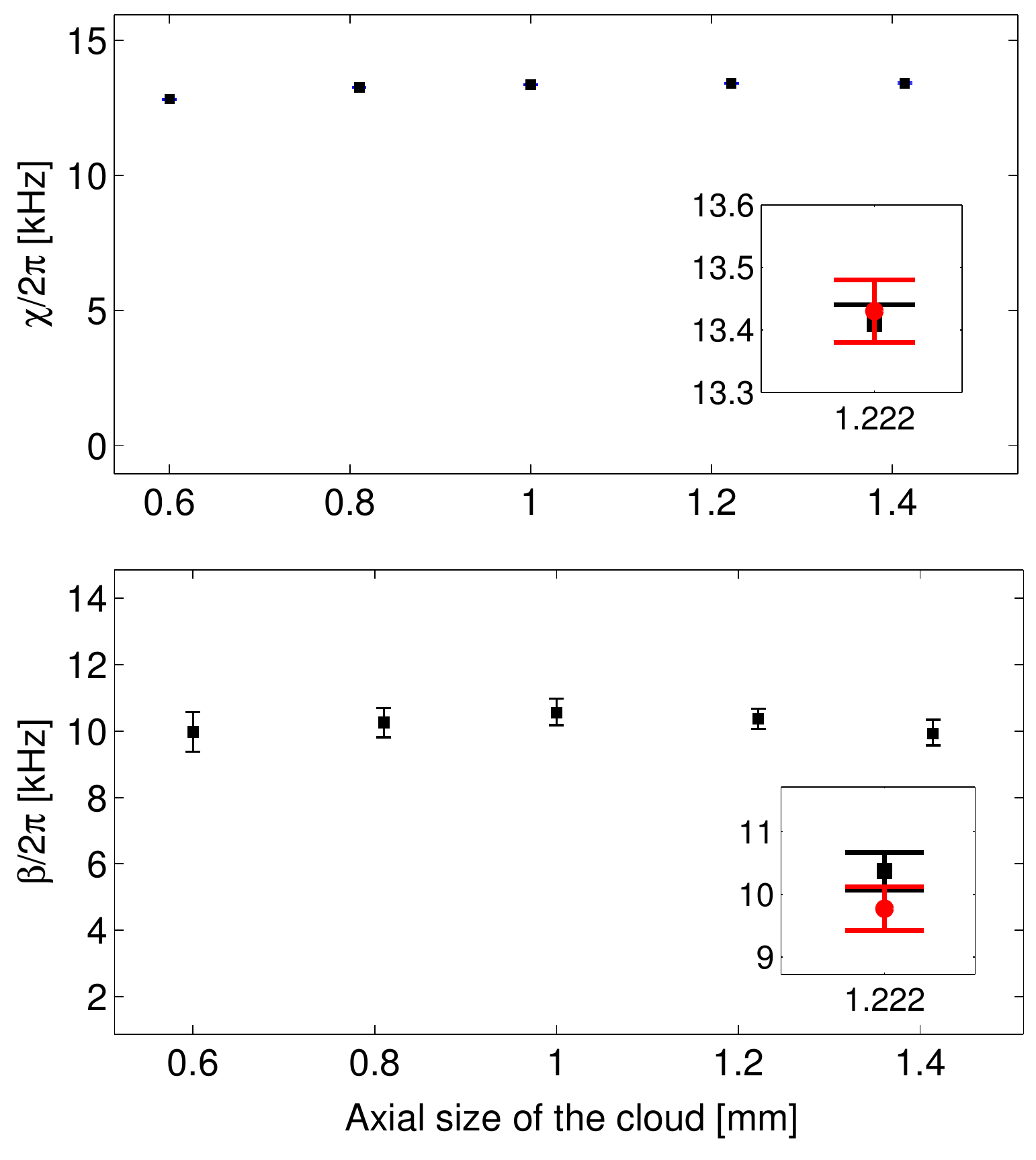}\\
  \caption{(Color online) Dispersive measurements of the on-resonance Rabi frequency (a) and the field gradient (b) for various axial sizes of the atomic sample. The values are obtained by fitting the model \eref{model2} to the dispersively obtained data. The errorbars show the 2$\sigma$ confidence intervals. Insets compare the values obtained using the dispersively ({\tiny$\blacksquare$}) and absorptively (\textcolor{red}{$\bullet$}) measured time evolution.}\label{figxy1}
\end{center}
\end{figure}

When fitting the model \eref{model2} to the absorptive data in \fref{fig:bloch1}(c), we obtain excellent agreement. From the fit we extract an on-resonance Rabi frequency of $\chi = 2\pi \times 13.43(2)$\,kHz and a field gradient of $\beta =2\pi \times 9.7(3)$\,kHz/mm, where the numbers in brackets are the $2\sigma$ confidence intervals.

Since the model \eref{model2} does not make any assumptions on how the population measurements are made, we can also apply it to the dispersive data presented in \fref{fig:disprecord}. Again, we obtain excellent agreement, as in the case for absorptive data.
The fitted $\chi$ and $\beta$ parameters of the model are plotted in \fref{figxy1}\,(a) and \fref{figxy1}\,(b), respectively, for five different spatial extents of the cloud in dipole trap. The values of the field gradient obtained in each case are in very close proximity to that obtained from the (time-consuming) destructive measurements (shown in the insets along with the corresponding values obtained dispersively) \footnote{The wider confidence interval for the first point in \fref{figxy1}(b) (which corresponds to a 10\,ms expansion time in the dipole trap after the IP trap switch-off) results from the instability of the field due to decaying eddy currents from the IP trap switch-off.}.

\subsection{Probe-induced decoherence and dephasing}\label{section:dephasing}
There are two important mechanisms through which the dispersive probe beam perturbs the coherent evolution of the system: the spontaneous photon scattering, and the differential light shift of the spin states involved. In contrast, the spontaneous photon scattering and the differential light shift caused by the dipole trapping beam are small and can be completely neglected for our beam parameters.

\subsubsection{Spontaneous scattering of probe photons}
The spontaneous photon scattering results from off-resonantly populating excited atomic states during the probe pulse \cite{Grimm2000,Petrov2007}. For our experimental parameters, we estimate the mean number of spontaneously scattered photons from a probe pulse (frequency triplet) to be about $9.5\times 10^{-5}$ per atom. This amounts to less than 0.1\% of the atoms scattering a photon over 100 dispersive pulses. In our experiment, we have about $3\times 10^{6}$ probe photons per pulse (about one photon per atom) in the red sideband of the probing triplet, and the time scale over which the decay due to spontaneously scattered photons becomes important is $\sim 50$~ms or longer.

\subsubsection{Dephasing model for a homogenous light shift}\label{section:dephasing_homogeneous}
\begin{figure}[tb!]
\flushright
  \includegraphics[width=\columnwidth]{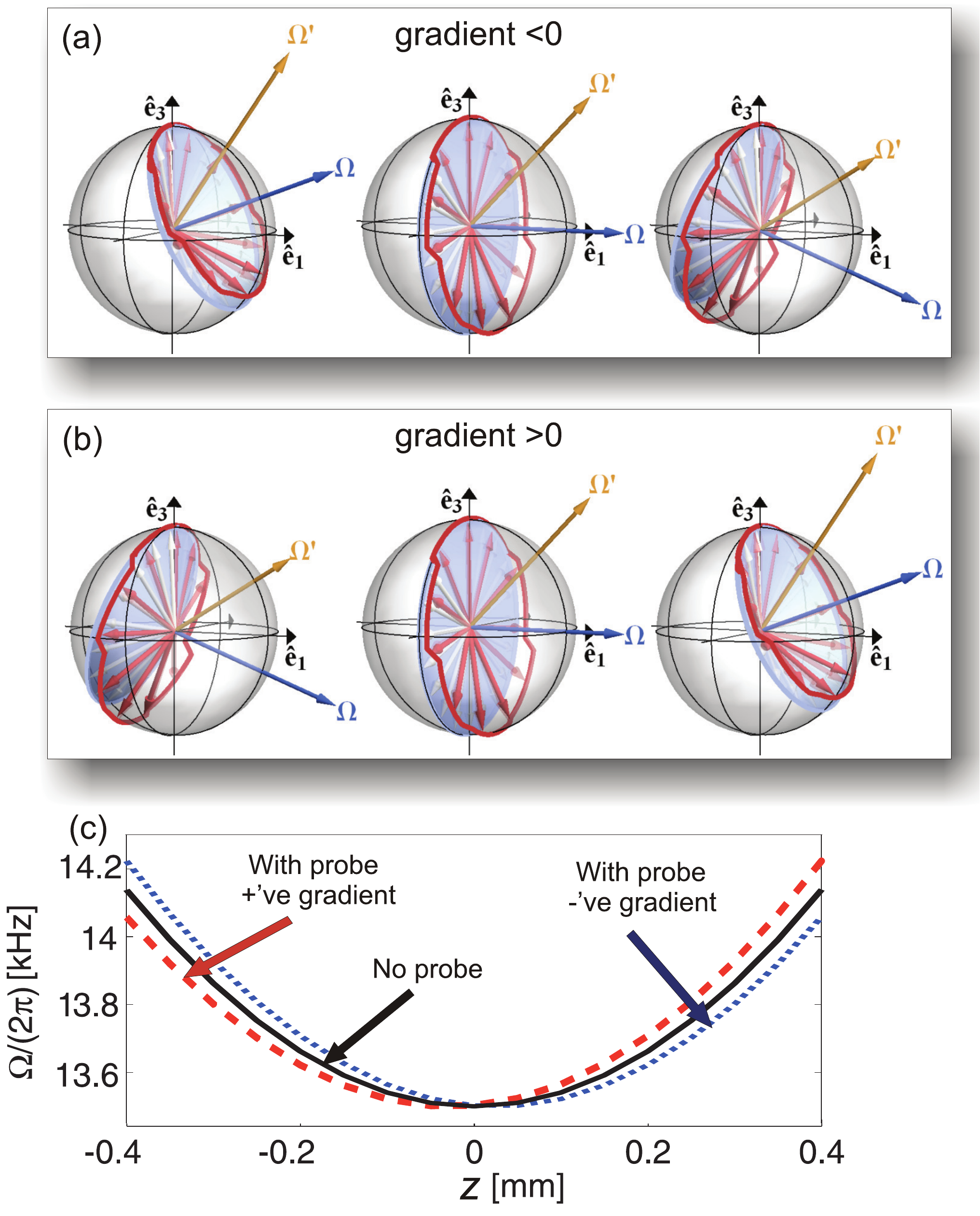}\\
  \caption{(Color online) Effect of a homogeneous differential light shift on the effective Rabi frequency. Bloch vector evolution along the sample for a positive (a) and a negative (b) field gradient. The gradient causes a position-dependent detuning away from the center (this detuning is assumed to be zero at $z=0$ in the absence of light). Without light, the Bloch vectors precess about $\bm{\Omega}(z)$ as encountered in \fref{fig:bloch1}(b). With light pulses applied periodically, precession about $\bm{\Omega}'(z)$ is interleaved into this evolution, effectively decrease the cycle time for $z<0$ ($z>0$) and a negative (positive) gradient. (c) The effective local Rabi frequency versus $z$ for the cases of a negative and a positive linear field gradient and a homogeneous light shift. Also shown is the hyperbola \eref{eqrabifreq} --- symmetric about $z=0$ --- that describes \emph{both} negative and positive field gradients in the absence of a light shift.}
  \label{fig:dephasing2}
\end{figure}

We quantify the differential light shift induced by the probe beam by the (local) dimensionless quantity $\rho ~=~ \xi t_{\mathrm{probe}}/\hbar$, where $\xi$ is the (local) differential light shift caused by the probe and $t_{\mathrm{probe}}$ is the duration of each probe pulse. For the parameters used in our experiments, the maximum differential light shift is $2\pi\times 34$\,kHz, and the corresponding $\rho_{\mathrm{max}} \simeq 0.1$. Due to the Gaussian intensity profile of the probe beam, the differential light shift varies across the atomic sample. The Rayleigh range is $\sim1.5$~cm for the probe beam, which is an order of magnitude higher the axial size of the atomic sample. This means that the light shift along this direction is essentially homogeneous.

In order to gain insight into the physical process, we again make use of the Bloch sphere picture. Locally, the dispersive atom-light interaction during a probe pulse causes a rotation of the Bloch vector with respect to the $\mathbf{\hat{e}}_3$-axis by an angle $\rho$. For blue-detuned probe light (as in our case), $\rho$ is positive. When applied during the coherent microwave evolution, a series of probe pulses effectively shifts the Bloch vector away from the circle on which its tip would otherwise lie, as illustrated in \fref{fig:dephasing2}(a) and in \fref{fig:dephasing2}(b), in the presence of negative and positive field gradients, respectively. Consequently, for a positive (negative) field gradient, the resonance moves to the left (right) along the cloud. This notion is reasserted in \fref{fig:dephasing2}(c) which shows the calculated local Rabi frequencies along the $z$-direction in the presence of light pulses for both positive and negative field gradients, obtained by numerically solving the Bloch equation, and assuming $\rho_{\mathrm{max}} = 0.1$ and $\chi = 2\pi \times 13.5$~kHz.
\begin{figure}[b!]
  \includegraphics[width=\columnwidth]{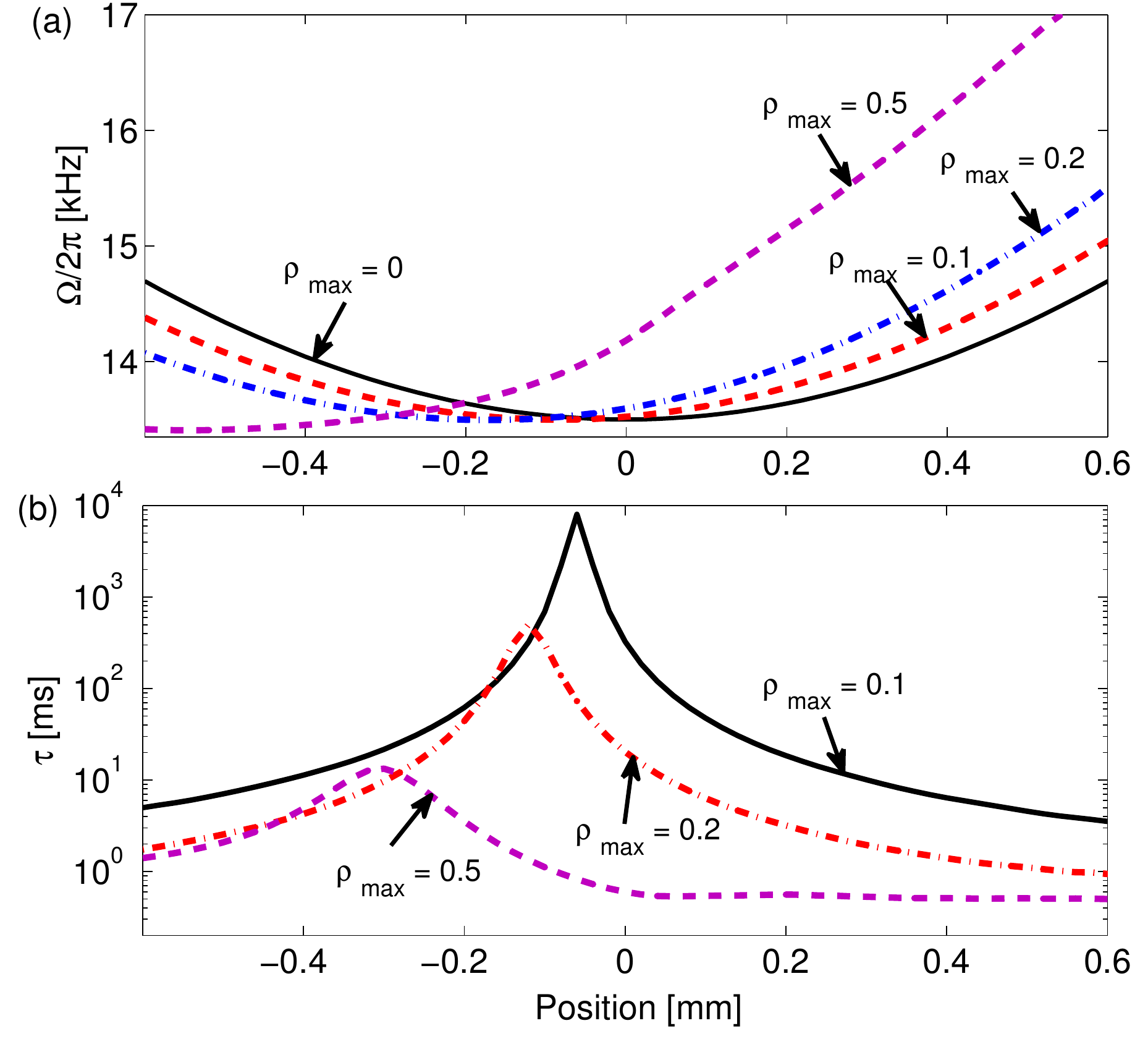}\\
  \caption{(Color online) (a) Effective Rabi frequency along the $z$-axis (axial direction of the atomic sample) in the presence of a positive magnetic field gradient. The symmetry encountered in the absence of a perturbing light shift ($\rho_{\rm max}=0$) is broken when a \emph{radially} inhomogeneous light field of increasing strength $\rho_{\rm max}=0.1,0.2,0.5$ is applied. In addition to a translation of the minimum also found in the homogenous case \fref{fig:dephasing2}(c), the inhomogeneity leads to a ``tilt''. (b) Characteristic dephasing decay time $\tau$ versus position. }
  \label{fig:dephasing3}
\end{figure}
\subsubsection{Dephasing model for an inhomogenous light shift}\label{section:dephasing_inhomogeneous}
The radial size of our atomic cloud ($\sim 30$\,$\mu$m) is comparable to the waist of the dispersive beam ($65$\,$\mu$m), which means that the considerations from \sref{section:dephasing_homogeneous} need to be extended to include the effect of inhomogeneous light shift in the radial direction. For a given axial position $z$, atoms at various radial distances experience different light shifts $\xi$, which results in radially dependent azimuthal phase shifts on the Bloch sphere during a probe pulse. This leads to \emph{local} dephasing at a given axial position.

The time evolution of the collective pseudospin vector $\overrightarrow{\mathcal{R}}$ for each point $z$ is the sum of the time evolution of the Bloch vector for each radial position $r$ weighted by $\bar{n}(z,r,t)$, the (normalized) local radial atomic density, $\overrightarrow{\mathcal{R}}(z,t)=\int d r\bar{n}(z,r)\mathbf{R}(z,r,t)2 \pi r$.  We numerically solved this equation for $\overrightarrow{\mathcal{R}}$ for different values of the peak differential light shift $\rho_{\rm max}$ while assuming a positive magnetic field gradient. We summarize the results in terms of the distributions of local effective Rabi frequency and the dephasing decay time in \fref{fig:dephasing3}. As is evident from \fref{fig:dephasing3}(a), the radially inhomogeneous light shift leads to a breaking of symmetry of the local Rabi frequency about the minimum point (a ``tilt''). It also shows that the collective pseudospin vector on the right (left) side of the cloud precesses faster (slower) in the presence of dispersive pulses causing faster (slower) spin dynamics on the right (left) side of the elongated atomic sample with respect to the unperturbed case. Figure \ref{fig:dephasing3}(b) shows that for the probe pulse parameters used in our experiments ($\rho_{\mathrm{max}} = 0.1$), the expected characteristic decay time $\tau$ owing to the dephasing caused by the inhomogeneous light shift is much longer than the time scale of the coherent manipulation and dispersive interrogation in our experiment (within the relevant length scale).
\begin{figure*}[bt!]
\begin{center}
  \includegraphics[width=\textwidth]{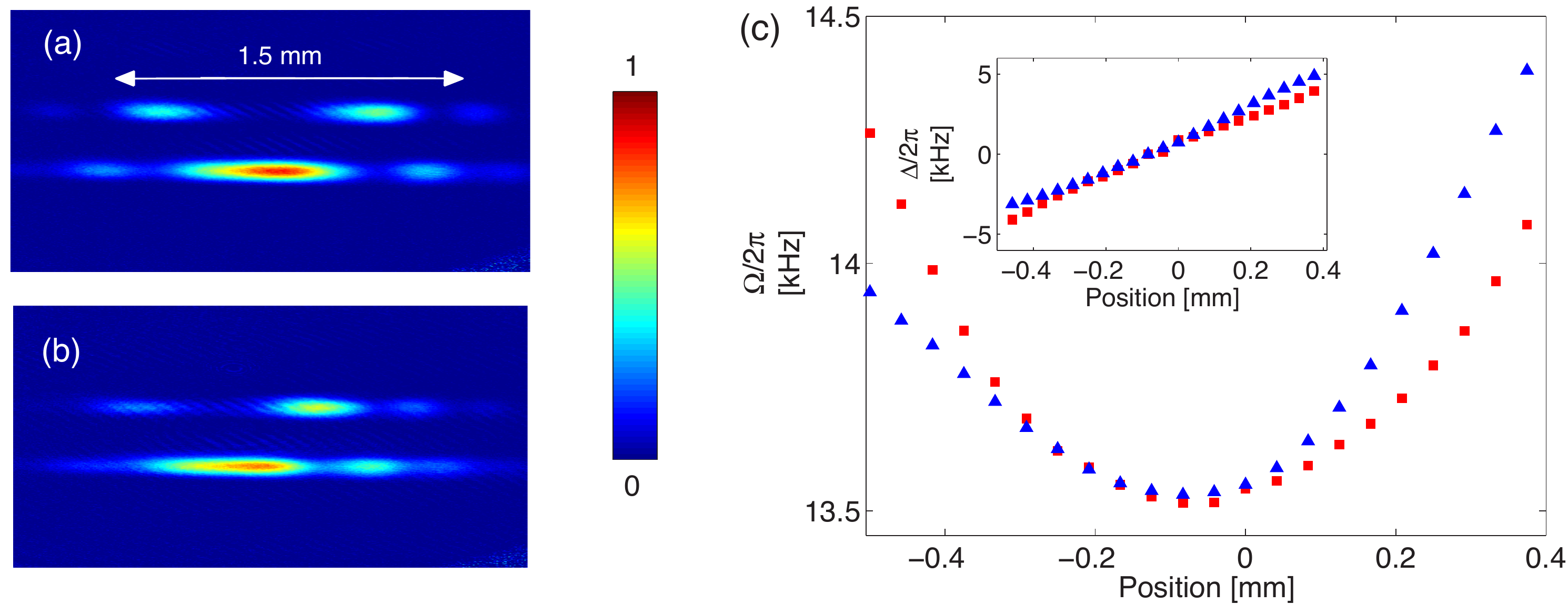}
  \caption{(Color online) State resolved absorption images taken after 4 full Rabi cycles without probe pulses (a) and with probe pulses (b). A series of such images were taken after exposing the atoms to microwave for various durations. (c) Local Rabi frequencies determined from the absorption images using the method discussed in the text for the case with no probing ({\tiny\textcolor{red}{$\blacksquare$}}) and the case with probe pulses on ({\textcolor{blue}{$\blacktriangle$}}). The ``tilt'' of the hyperbola from right to left indicates that the associated detuning has a positive gradient, as explained in the text. The inset shows the local detunings for the two cases as inferred from \eref{eqrabifreq}.}
   \label{figxy2}
\end{center}
\end{figure*}

\subsubsection{Experiment: symmetry breaking from a light shift}\label{section:ex_local_rabi}
Finally, we make a local analysis on the series of state resolved absorption images taken after various microwave exposure times in the absence and presence of dispersive probe pulses. \Fref{figxy2}(a) and \fref{figxy2}(b) show the absorption images taken at $4 \times T_{R}$ in the two cases, respectively, revealing a clear breaking of symmetry for the latter case. We divide the time series of absorption images in the two cases into equal sized bins and count the number of atoms in each state residing in a bin. From this the time evolution of the population in each bin is constructed and we extract the local (effective) Rabi frequency for each bin. The results are presented in \fref{figxy2}(c) for the case of a $40~\mu$m bin size. Evidently, dispersive probe pulses cause a small shift of the minima and break the symmetry of the curve, a result of the accompanying differential light shift of the transition, as discussed in \sref{section:dephasing_homogeneous} and \sref{section:dephasing_inhomogeneous}. In particular, a behavior akin to that predicted by \fref{fig:dephasing3}(a) is encountered. The inset of \fref{figxy2}(c) shows the corresponding local detuning $\Delta(z) = [\Omega(z)^2 - \chi^2]^{1/2}$, where $\chi$ is the minimum of $\Omega(z)$ for the two cases. The sign of the slope of $\Omega(z)$ is determined to be \emph{positive} based on the directions of the ``shift'' and the ``tilt'' of the hyperbolic shape of \fref{figxy2}(c) for the dispersively probed case following the reasoning of \sref{section:dephasing_inhomogeneous}. The values of $\beta$ obtained from the corresponding linear fits [$\beta = 2\pi \times 10.26(43)$~kHz/mm for the unperturbed case and $\beta = 2\pi\times10.24(42)$~kHz/mm for the dispersively probed case] are in very good agreement with each other and with the values obtained in the two sample-integrated procedures (dispersive and absorptive) presented in \sref{section:gradient_magnetimetry}.
\section{Conclusion}\label{section:conclusion}
In this paper we have investigated a heterodyne, dispersive optical probing scheme
applied to elongated ultracold rubidium clouds residing in an inhomogeneous magnetic field. A Zeeman shift gives rise to axial dephasing when driving near-resonant Rabi oscillations between two
hyperfine states. With the dispersive probe, the resulting decay in net polarization/magnetization of the sample could be followed in ``real time''. In particular, the dispersively observed non-exponential washing out of Rabi oscillations was described well by a simple model from which a value of the magnetic field gradient could be extracted. This value for the magnetic field gradient was in excellent correspondence with complementary measurements based on absorption images. For our demonstration we applied a field gradient of about 500~nT/mm, leading to a center-to-edge change
in Rabi frequency of $\sim4\%$ from the $\sim13$~kHz central and on-resonance Rabi frequency. Fitting our model function over a time interval $\sim1$~ms (yielding more than 10 cycles) allows us to
determine this gradient to within 25~nT. This underscores the potential to perform gradient magnetometry to this accuracy with a $\gtrsim$kilohertz bandwidth using the dispersive probing approach based on a single experimental run \footnote{For the dispersive data analyzed in this paper we chose to consider averages of five consecutive experimental runs to boost signal to noise ratio. The single-run signal to noise ratio could however be improved straightforwardly by employing a probe triplet with a stronger carrier further detuned away from atomic resonance.}.

The absorption images of prolate clouds at the end of the Rabi drive revealed the formation of (continuous) magnetic order along the sample similar to previously reported findings \cite{Matthews1999,Hamner2013}, but also how the small but finite light shift from the dispersive probe breaks the symmetry of the spin patterns. We used this to the effect of determining the sign of the applied magnetic field gradient. As outlined in \sref{section:backgroundb} there is significant interest in spin dynamics of ultracold atomic systems, and we speculate if this mechanism can be employed for spin wave engineering similar to \cite{McGuirk2011}.

We finally point towards possible future applications of the experimental scheme outlined in this paper, in the context of probing microscopic structures. For example, atoms have recently been used as a scanning ``soft'' probe for measuring the height and position of free-standing carbon nanotubes \cite{Gierling2011}. Experiments of this kind have so far relied on absorption imaging for measuring the atomic loss constituting the meter variable. Dispersive probing strategies hold the promise of vastly improving the data acquisition rate over such destructive detection schemes. Furthermore, atomic clouds have been envisaged for sensing the magnetic field from oscillating micro-mechanical wires carrying a current \cite{Kalman2012}. We believe a Rabi driven, dispersively probed, prolate atomic cloud would form an interesting interface to such systems.

\acknowledgments
We thank Ana Rakonjac and Sascha Hoinka for their contribution to the construction of the trap setup and Thomas McKellar for his role in developing the dispersive probing scheme. We are grateful to Peter Engels, who made results of his group on phase winding in a field gradient available to us, and to Andrea Bertoldi for generously surrendering some spare high speed photo diodes to us. Finally, NK acknowledges motivating discussions on spin waves and spinor systems with Blair Blakie.

This work was supported by FRST contract NERF-UOOX0703 and the Marsden Fund of New Zealand (Contract No. UOO1121).
BJS was supported by a University of Otago Postgraduate Publishing Bursary.

\end{document}